\documentclass[twocolumn]{aastex63}

\newcommand\aastex{AAS\TeX}

\usepackage{tcolorbox}
\usepackage{savesym}
\usepackage{soul}

\usepackage{amsmath}
\usepackage{capt-of}
\usepackage{graphicx}
\usepackage{lineno}
\usepackage[T1]{fontenc}
\usepackage{afterpage}
\usepackage{soul}
\savesymbol{tablenum}
\usepackage{verbatim}
\usepackage{siunitx}
\restoresymbol{SIX}{tablenum}
\DeclareUnicodeCharacter{2212}{-}
\definecolor{orcidlogocol}{HTML}{A6CE39}

\makeatletter

\patchcmd{\NAT@citex}
  {\@citea\NAT@hyper@{%
     \NAT@nmfmt{\NAT@nm}%
     \hyper@natlinkbreak{\NAT@aysep\NAT@spacechar}{\@citeb\@extra@b@citeb}%
     \NAT@date}}
  {\@citea\NAT@nmfmt{\NAT@nm}%
   \NAT@aysep\NAT@spacechar\NAT@hyper@{\NAT@date}}{}{}

\patchcmd{\NAT@citex}
  {\@citea\NAT@hyper@{%
     \NAT@nmfmt{\NAT@nm}%
     \hyper@natlinkbreak{\NAT@spacechar\NAT@@open\if*#1*\else#1\NAT@spacechar\fi}%
       {\@citeb\@extra@b@citeb}%
     \NAT@date}}
  {\@citea\NAT@nmfmt{\NAT@nm}%
   \NAT@spacechar\NAT@@open\if*#1*\else#1\NAT@spacechar\fi\NAT@hyper@{\NAT@date}}
  {}{}

\makeatother

\shorttitle{\aastex\ OGLE-2017-BLG-1434}
\shortauthors{Blackman et al.}

\begin{document}

\title{OGLE-2017-BLG-1434Lb: Confirmation of a Cold Super-Earth using Keck Adaptive Optics} 

\author[0000-0001-5860-1157]{J.W. Blackman}
\affil{School of Natural Sciences, University of Tasmania,
Private Bag 37 Hobart, Tasmania 7001 Australia}
\affil{Sorbonne Universit\'es, UPMC Universit\'e Paris 6 et CNRS, 
UMR 7095, Institut d'Astrophysique de Paris, 98 bis bd Arago,
75014 Paris, France} 

\author[0000-0003-0014-3354]{J.-P. Beaulieu}
\affil{School of Natural Sciences, University of Tasmania,
Private Bag 37 Hobart, Tasmania 7001 Australia}
\affil{Sorbonne Universit\'es, UPMC Universit\'e Paris 6 et CNRS, 
UMR 7095, Institut d'Astrophysique de Paris, 98 bis bd Arago,
75014 Paris, France}

\author[0000-0003-0303-3855]{A.A. Cole}
\affil{School of Natural Sciences, University of Tasmania,
Private Bag 37 Hobart, Tasmania 7001 Australia}

\author{N. Koshimoto}
\affil{Laboratory for Exoplanets and Stellar Astrophysics, NASA/Goddard Space Flight Center, Greenbelt, MD 20771, USA}
\affil{Department of Astronomy, Graduate School of Science, The University of Tokyo, 7-3-1 Hongo, Bunkyo-ku, Tokyo 113-0033, Japan}

\author[0000-0002-9881-4760]{A. Vandorou}
\affil{School of Natural Sciences, University of Tasmania,
Private Bag 37 Hobart, Tasmania 7001 Australia}



\author{A. Bhattacharya}
\affil{Laboratory for Exoplanets and Stellar Astrophysics, NASA/Goddard Space Flight Center, Greenbelt, MD 20771, USA}

\author[0000-0002-7901-7213]{J.-B. Marquette}
\altaffiliation{Associated with Sorbonne Universit\'es, UPMC Universit\'e Paris 6 et CNRS, 
UMR 7095, Institut d'Astrophysique de Paris, 98 bis bd Arago,
75014 Paris, France}
\affil{Laboratoire d'astrophysique de Bordeaux, Univ. Bordeaux, CNRS, B18N, all\'ee Geoffroy Saint-Hilaire, 33615 Pessac, France}

\author[0000-0001-8043-8413]{D.P. Bennett}
\affil{Laboratory for Exoplanets and Stellar Astrophysics, NASA/Goddard Space Flight Center, Greenbelt, MD 20771, USA}


\begin{abstract}
The microlensing event OGLE-2017-BLG-1434 features a cold super-Earth planet which is one of eleven microlensing planets with a planet-host star mass ratio $q < 1 \times 10^{-4}$. We provide an additional mass-distance constraint on the lens host using near-infrared adaptive optics photometry from Keck/NIRC2. We are able to determine a flux excess of $K_L = 16.96 \pm 0.11$ which most likely comes entirely from the lens star. Combining this with constraints from the large Einstein ring radius, $\theta_E=1.40 \pm 0.09\;\si{mas}$ and OGLE parallax we confirm this event as a super-Earth with mass $m_p = 4.43 \pm 0.25M_\Earth$. This system lies at a distance of $D_L = 0.86 \pm 0.05\,\si{kpc}$ from Earth and the lens star has a mass of $M_L=0.234\pm0.012M_\odot$. We confirm that with a star-planet mass ratio of $q=0.57 \times 10^{-4}$, OGLE-2017-BLG-1434 lies near the inflexion point of the planet-host mass-ratio power law.
\end{abstract}



\section{Introduction} \label{sec:intro}
The core accretion theory of planet formation \citep{Pollack1996} predicts a planetary desert at intermediate planet/host mass ratios of $1 < q/10^{-4} < 4$ \citep{Suzuki2016}.  This is due to runaway gas accretion which involves the rapid accumulation of hydrogen and helium gas onto protoplanetary cores as they reach masses of $\sim10M_\Earth$. This results in a dearth of intermediate-mass giant planets between Saturn mass planets at $\sim$95$M_\Earth$ and failed gas giants at $\sim$10$M_\Earth$. The predicted lack of planets with these masses is however in conflict with planet demographics determined from microlensing observations \citep{Suzuki2016, Suzuki2018}.\\ 
\indent \cite{Suzuki2016} compiled a sample of 30 planetary microlensing events from the MOA (Microlensing Observations in Astrophysics) survey observed between 2007 and 2012 and compared that with population synthesis models. They show that these models underestimate the number of planets with $1 < q/10^{-4} < 4$ by a factor of ten. They estimate the mass ratio distribution of exoplanets to be a broken power law with a break occurring at $q_{br}=1.7 \times 10^{-4}$. A subsequent study used 15 low mass-ratio microlensing events and estimated this break to instead be at $q_{br}\sim0.55 \times 10^{-4}$ \citep{Jung2019}. While the distribution at higher mass ratios is a well-sampled as a decreasing power-law, there are only eleven planets with mass ratios with  $q < \times 10^{-4}$, which is why arriving at robust statistical conclusions in this region of parameter space is difficult.\\  \indent In this paper we discuss the planetary microlensing event OGLE-2017-BLG-1434 \citep{Udalski2018}, a cold super Earth with mass ratio $q=0.57 \times 10^{-4}$. 
A measurement of the microlens parallax meant that the physical parameters are comparatively well measured, with the mass of the planet and its stellar host determined to be $m_p = 4.4 \pm 0.5M_\Earth$ and $M_L=0.23\pm0.03M_\odot$, respectively. However, as is typical in microlensing event analysis, a Bayesian analysis with constraints from the Einstein radius crossing time and measurable secondary effects (for example, due to the finite size of the source) are required in combination with a galactic model \citep{Sumi2011, Bennett2014a} to obtain estimates of these physical parameters. Without secondary constraints these estimates usually have an accuracy of 30-40\%.\\
\indent The first of these constraints can be derived from the sharp light curve features of many binary microlensing events. This enables one to measure the finite angular source radius \citep{Beaulieu2018a} and hence a relationship between the lens mass and distance: 
\begin{equation}
\begin{aligned}
M_L=\frac{\theta_E^2}{\kappa\pi_{rel}}
\label{eq:md1}
\end{aligned}
\end{equation}
where $\theta_E$ is the Einstein radius, $\pi_{rel}$ the relative parallax and $\kappa= 4G/{c^2\mathrm{AU}=8.144\;\si{mas}\;M_\odot^{-1}}$.
A second constraint on the lens mass and distance can be found by measuring the microlensing parallax, $\pi_E$. This can be determined via detection of the Earth's orbital motion or between two spatially separated observing sites:\\
\begin{equation}
\begin{aligned}
M_L=\frac{\pi_{rel}}{\kappa\pi_{\mathrm{E}}^2}
\label{eq:md2}
\end{aligned}
\end{equation}
Space-based parallax from observatories such as Spitzer are ideal in constraining $\pi_E$, though systematic errors may be a problem for both orbital and space-based parallax \citep{Koshimoto2020a, Gould2020, Dang2020, Poleski2016}. Finally, a third mass-distance relation can be obtained by measuring the flux of the lensing system using high-angular resolution observations from 8-10 metre class telescopes. This makes it possible to decouple the source and lens contributions from that of other blend stars. The relation that follows is:
\begin{multline}
m_L(\lambda)=10+5log(D_L/1\:\si{kpc})+A_L(\lambda)\\+M_{iso}(\lambda,M_L,age,[Fe/H])
\label{eq:md3}
\end{multline}
where $m_L(\lambda)$ is the apparent magnitude of the lens, $D_L$ is the distance to the lens, $A_L(\lambda)$ is the extinction to the lens in band $\lambda$, and $M_{iso}$ is an absolute magnitude derived from stellar isochrones. It is this relationship which we use in this paper to better quantify the mass and distance of OGLE-2017-BLG-1434.\\
\begin{figure*}[th!]
\centering
\includegraphics[width=1.0\textwidth]{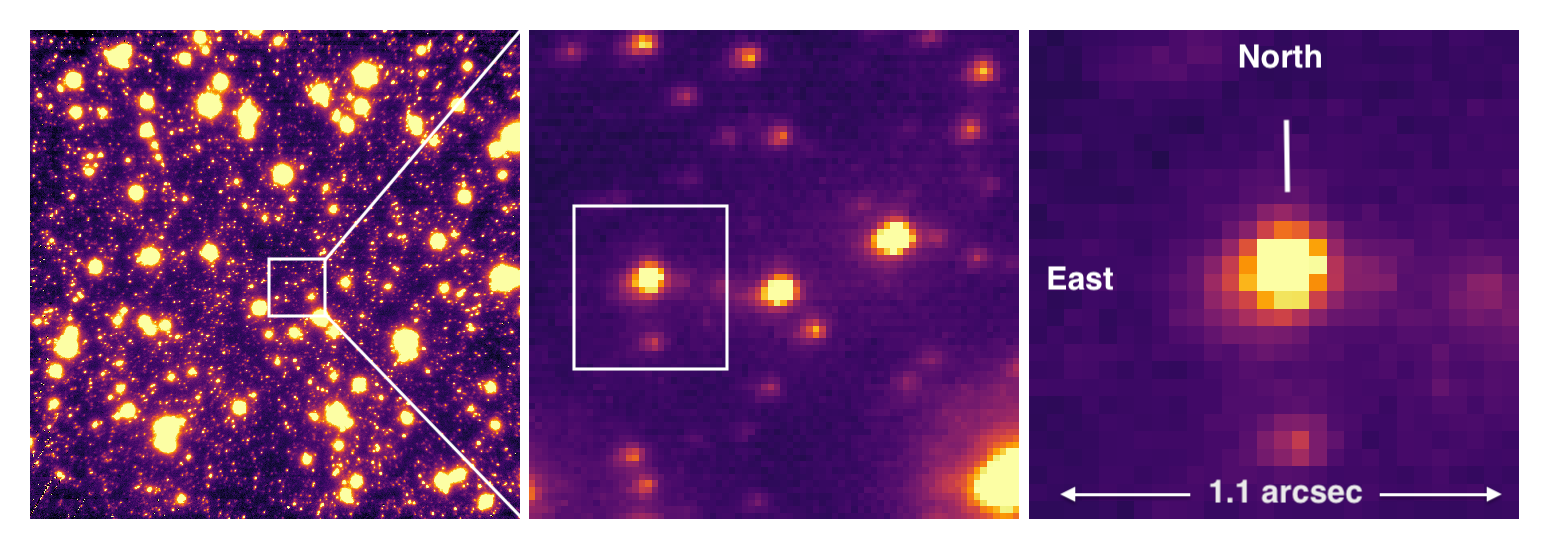}
\caption{K-band images of OGLE-2017-BLG-1434 obtained on August 7, 2018 with Keck/NIRC2. On the far left is an image taken with the wide camera with a 40 arcsec field of view. On the right are two enlargements of this frame showing the position of the source/lens blend, indicated with a white line.}
\label{fig:keck}
\end{figure*}
\indent We supplement the  discovery  paper of this event with  the  addition  of  high-angular resolution data obtained using the NIRC2 imager on the Keck II telescope on Mauna Kea, Hawaii. We obtained follow-up photometry in August 2018 as part of a NASA Keck Key Strategic Mission Support (KSMS) proposal in support of the Nancy Grace Roman Space Telescope (formerly WFIRST, \cite{Spergel2015}). This program is designed to acquire adaptive optics data of microlensing events in order to constrain the lens flux and/or the source-lens relative proper motion, and hence determine accurate estimates of the planet and host masses. Results from this program can be found in a number of studies \citep{Beaulieu2016, Bhattacharya2018, Beaulieu2018, Vandorou2020, Bennett2020, Bhattacharya2020, Terry2021}. We use this photometry to measure the lens apparent magnitude $m_L(\lambda)$ from Eq. \ref{eq:md3} and hence determine a relationship between $D_L$ and $M_L$. The finite source size, ground parallax and this newly obtained adaptive optics data, combined with theoretical isochrones, allows us to refine the estimate of the mass-distance relationship and confirm this event as a cold super Earth, $m_p = 4.43 \pm 0.25M_\Earth$, very nearby for a microlensing planet at a distance of $D_L = 0.86 \pm 0.05\;\si{kpc}$.

\section{The cold super-Earth OGLE-2017-BLG-1434} \label{sec:ao}

OGLE-2017-BLG-1434 was detected by the OGLE Early Warning System at UT19:33 on 25 Jul 2017 using the OGLE 1.3m telescope at Las Campanas. Located at $(RA,Dec)_{J2000} = (17^h53^m07s.29^s,−30^{\circ}14'44.6'')$, complementary follow-up data was taken with telescopes from the Korea Microlensing Telescope Network (KMTNet, \cite{Kim2016}) and the MiNDSTEp collaboration at the 1.54m Danish Telescope at La Silla, Chile. The majority of the observations were taken in I-band with V-band data obtained to determine source colors. The resultant light curve was similar to that of a traditional point-lens with the addition of deviations near the event peak. \cite{Udalski2018} determined the source brightness from the fitted light curve and used red clump extinction estimates from \cite{Bensby2013} and \cite{Nataf2013} to derive an angular source size of $\theta_*=0.657 \pm 0.041 \;\si{\mu as}$. This results in an Einstein ring radius of $\theta_E = 1.40 \pm 0.09\;\si{mas}$ and a relatively large lens-source relative proper motion, $\mu = 8.1 \pm 0.5\;\si{mas}\;\si{yr^{-1}}$. This large Einstein ring radius indicates that the lens must either be very close or very massive. The long timescale of the event ($t_E \sim 63$ days) suggests the presence of microlens parallax, though \cite{Udalski2018}'s models were unable to decouple parallax from orbital motion.  \cite{Udalski2018} finds two degenerate ($\pm u_0$) models with parameters which are statistically indistinguishable within $1\sigma$.\\
\indent With a year having elapsed since the peak magnification of OGLE-2017-BLG-1434, the relative source-lens proper motion is such that we would not expect to be able to resolve these blended components with the best achievable resolution of Keck/NIRC2 of $\sim50\:\si{mas}$. We can, however, compare the predicted source magnitude from the light-curve model with the measured flux of the object at the position of the source, and attempt to decouple and characterize the excess flux.
\newpage
\subsection{Keck Observations}

We observed OGLE-2017-BLG-1434 with Keck/NIRC2 on August 7, 2018 (HJD: 2458337.80080). 29 images were obtained in the short $K_s$ band using the wide camera. For simplicity in the rest of this paper we will drop the subscript and refer to the magnitudes simply as $K$. The wide camera has a plate scale of 0.03968 arcsec pixel$^{-1}$ and a field of view of 40 arcseconds. The best 15 of these images were stacked using \texttt{SWARP} \citep{Bertin2010} and calibration performed by cross-matching with the VVV catalog, following the process described in \cite{Blackman2020} and \cite{Vandorou2020}. The mean full-width half-maximum (FWHM) is 120 mas along the North axis and 90 mas along the east, indicating an elongation of the point spread function (PSF). This elongation is not severe enough to affect our photometry. The resulting stacked image can be seen in the left panel of Fig. \ref{fig:keck}. To determine the location of the source we use precise astrometry determined from the OGLE-III survey image. The OGLE image
coordinates measured during magnification were (X,Y) = (1829.25, 3196.66) (Private communication, Udalski, 5/12/18) with the source having an OGLE ID of 198963. This places the object at $(RA,Dec)_{J2000} = (17^h53^m07s.312^s,−30^{\circ}14'44.37'')$ in our stacked image (Fig. \ref{fig:keck}). Photometry was performed on this image using \texttt{SExtractor} \citep{Bertin1996}. We find the magnitude of the object at the location of the source to be:
\begin{align}
K_{blend} = 16.61 \pm 0.07
\label{eq:measuredflux}
\end{align}
where the blend is defined here as the total flux of the object.
\subsection{Extinction Estimates}
\begin{figure}[th]
\centering
\includegraphics[width=0.47\textwidth]{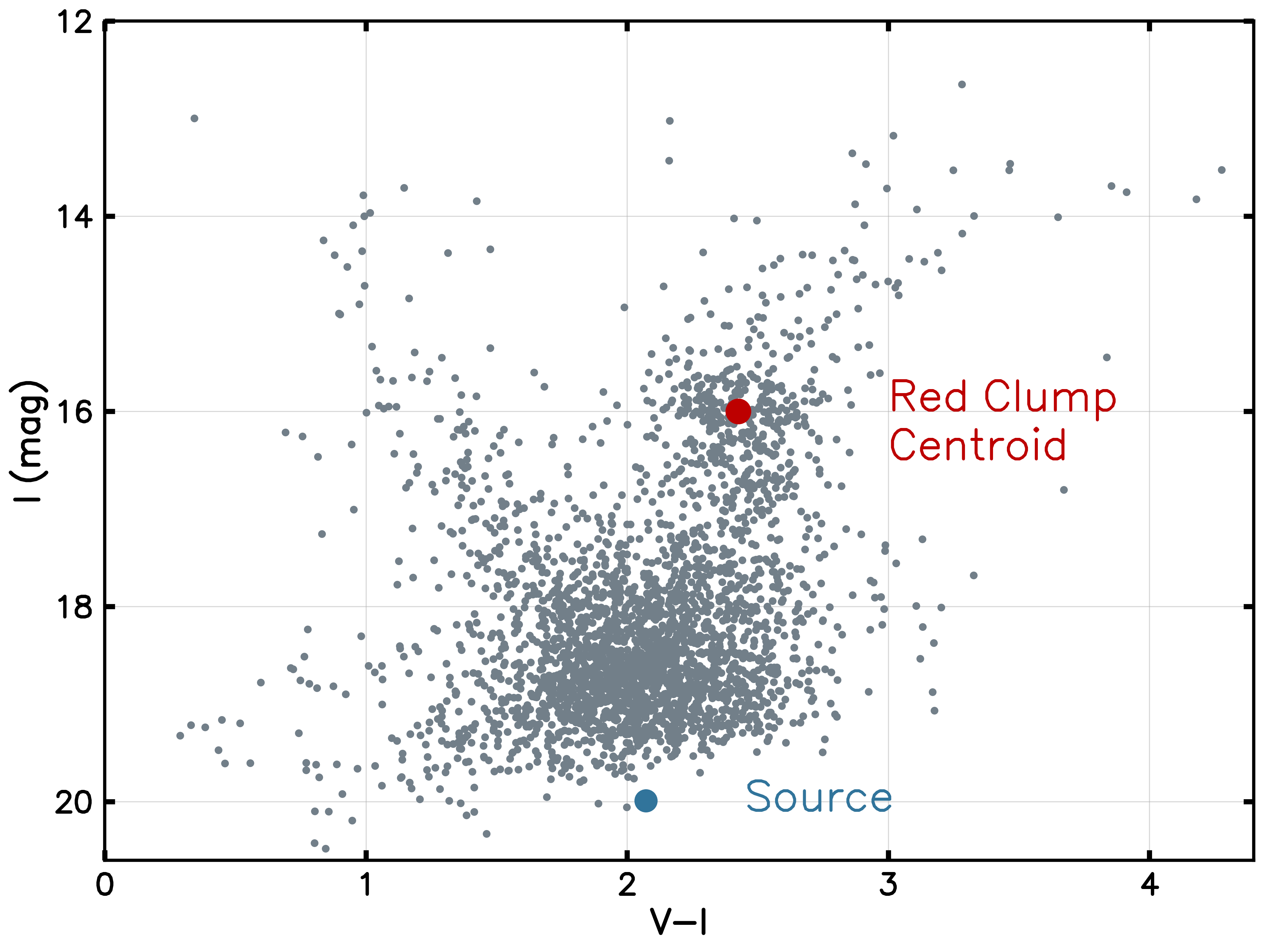}
\caption{OGLE-III calibrated color-magnitude diagram in V and I. The red circle marks the centroid of the red giant clump while the blue circle indicates the position of the source. The centroid of the red clump is located at $[(V-I)_{RC},I_{RC}] = (2.41, 16.09)$.}
\label{fig:cmd}
\setlength{\belowcaptionskip}{-18pt}
\end{figure}

\cite{Udalski2018} determines the intrinsic source color and magnitude to be: $[(V-I)_{S0},I_{S0}] = (0.732, 18.45) \pm (0.025, 0.063).$
Following the color relations of \cite{Bessell1988} we can interpolate to find the V-K color, $(V-K)_{S0} = 1.59^{+0.05}_{-0.06}$, and hence the predicted intrinsic K-magnitude of the source, $K_{S0} = 17.59 \pm 0.09$. In order to compare this with our measured K magnitude, we must re-redden this using an estimate of the extinction on the path towards the source. We find $A_K=0.259\pm0.021$ using the OGLE extinction calculator\footnote{http://ogle.astrouw.edu.pl/cgi-ogle/getext.py} to estimate the K-band extinction at galactic coordinates $(l,b)=(0.28,2.07)$. This value is derived from a natural neighbour interpolation of good points from Table 3 in \cite{Nataf2013} and assuming $E(J-K_s)$ measurements from \cite{Gonzalez2012}. We use the values $E(V-I)=1.521\pm 0.125$, $R_{JKVI}=0.3195$ and the extinction law from \cite{Nishiyama2009} with the relationship $E(J-K_s) = R_{JKVI} E(V-I)$. The I-band extinction towards this part of the galactic bulge is estimated as $A_I=1.801.$\\
\indent To check our $A_K$ estimate we calculate the extinction directly from the OGLE-III field (Fig. \ref{fig:cmd}). Comparing this to the intrinsic brightness of the red clump, $[(V-I)_{RC0},I_{RC0}] = (1.06, 14.46)$ \citep{Nataf2013} gives $E(V-I)=1.55$ which is well within the error given by the OGLE extinction calculator. Using the value of $A_K=0.259\pm0.021$ we hence find a predicted source magnitude of $K_{\mathrm{predict}} = 17.85 \pm 0.09$. Subtracting this from Eq. \ref{eq:measuredflux} we find an excess flux of
\begin{align}
K_{\mathrm{excess}} = 17.03 \pm 0.11
\label{eq:exces}
\end{align}
We now re-correct for extinction but now only to the distance of the lens. The lens in \cite{Udalski2018} is predicted to be at a $0.86\pm 0.09\:\si{kpc}$, in front of more than half of the extinction along the 8 kpc line of sight towards the galactic bulge. We follow the procedure as in \cite{Bennett2015} and \cite{Beaulieu2018}, using the relationship
\begin{align}
A_{\mathrm{K}_{\mathrm{L}}}=(1-e^{-D_{\mathrm{L}/\tau_{\mathrm{dust}}}})/(1-e^{-D_{\mathrm{S}/\tau_{\mathrm{dust}}}})A_{\mathrm{K}_{\mathrm{S}}}
\end{align}
where the scale height of the dust towards the galactic bulge is $\tau_{dust} = (0.10 \pm 0.02\:\si{kpc})\sin b$ and $b=2.07$ the galactic longitude. In our case we calculate $A_{\mathrm{K}_{\mathrm{L}}}=0.073$ adopting a source distance of $D_S = 8.0 \pm 0.5\:\si{kpc}$ as predicted by the OGLE extinction calculator. If the light from this excess is entirely from the lens, or from a combination of objects at the same distance of $0.86\:\si{kpc}$, we find a excess flux of
\begin{align}
K_{\mathrm{0,excess}} = 16.96 \pm 0.11
\label{eq:blend}
\end{align}
\begin{figure*}[]
\centering
\includegraphics[width=0.8\textwidth]{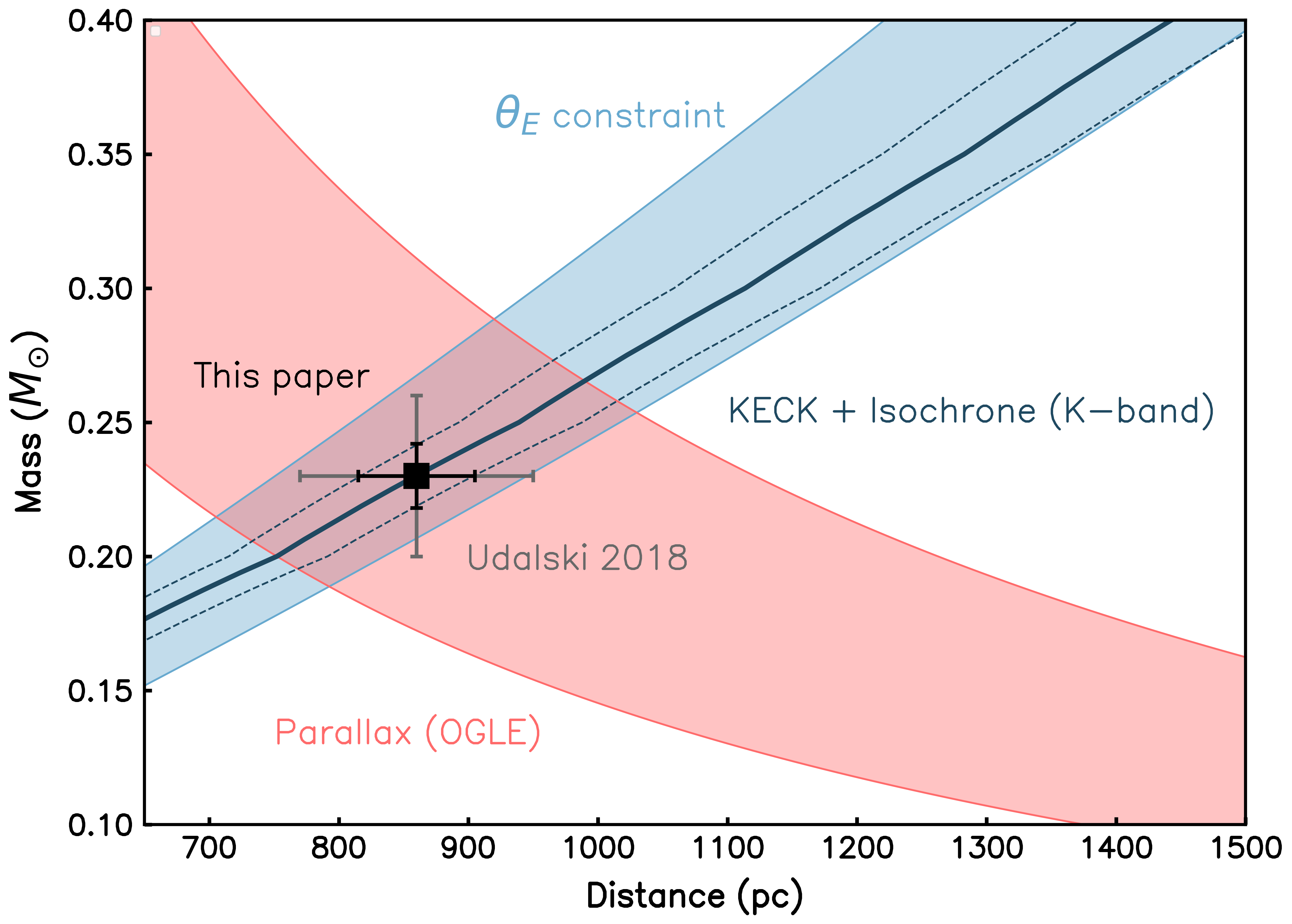}
\caption{Mass-distance diagram showing constraints from the Einstein Ring Radius ($\theta_E$, in blue), from OGLE parallax (in red) and our Keck measurement and K-band isochrones  \citep{Girardi2002}, the solid and dotted blue lines. The original estimate from \cite{Udalski2018} is shown with the grey cross at the intersection of the $\theta_E$ and OGLE parallax constraints. This paper adds the constraint from lens flux + isochrones. Our estimate is plotted as the black cross.}
\label{fig:md}
\end{figure*}
\begin{deluxetable*}{c @{\extracolsep{\fill}} cccc}
\tablecaption{Physical Parameters of OGLE-2017-BLG-1434\label{tab:results}}
\tablehead{\colhead{Parameter}& \colhead{Value}& with Keck AO & \cite{Udalski2018}}
\startdata
Lens Distance&$D_L$ (kpc)  & $0.86\pm0.05$& $0.86\pm0.09$  \\
Lens Mass&$M_L (M_\odot)$                  &  $0.234 \pm 0.012$& $0.234\pm0.026$ \\
Planet Mass&$m_p (M_\Earth)$                & $4.43\pm 0.25$& $4.4\pm0.5$ \\
Instantaneous 2D star-planet separation&$a_\perp (\mathrm{AU})$   & $1.18 \pm 0.10$& $1.18\pm0.14$ \\
\enddata
\end{deluxetable*}
\subsection{Lens Properties and Bayesian analysis} \label{sec:bayes}
We detect an excess flux aligned to the source to better than the best 90mas FWHM of our final swarped Keck image. In order to determine whether this light is (a) entirely from the lens, (b) from a companion to the lens, (c) a companion to the source, or (d) from an ambient star unrelated to either the lens or the source, we plot a mass-distance diagram assuming all the light is from the lens. Fig. \ref{fig:md} combines constraints from the Einstein Ring Radius, $\theta_E$, OGLE parallax and our flux measurement of $K_L = 16.96 \pm 0.11$ combined with theoretical isochrones or, in other words, Eq. \ref{eq:md1}, \ref{eq:md2} and \ref{eq:md3}. We find values for the lens mass and distance consistent with that determined by \cite{Udalski2018}, but with smaller uncertainties. The agreement between the model and our additional lens flux constraints is such that this excess flux is most likely entirely from the lens. \\
\indent To test this we perform a Bayesian analysis as described in \cite{Koshimoto2020}. Using the galactic model prior from \cite{Bennett2014a} and constraints from the large $\theta_E$, the observed $t_E$ and the measured flux excess of $K=16.61\pm0.07$, we find the excess to very likely be from the lens with a probability of 0.96 (Fig. \ref{fig:bayesian}a). In this calculation we deliberately exclude priors from parallax $\boldsymbol{\pi}_{\mathrm{E}}=\left(\pi_{\mathrm{E}, N}, \pi_{\mathrm{E}, E}\right)$. In this case the $\pi_{E,N}$ determined by the \cite{Udalski2018} models lie in the $3\sigma$ boundary while $\pi_{E,E}$ is slightly more likely and sits within the $2\sigma$. When using the $\theta_E$ and $t_E$ constraints only without our Keck measurements, the large parallax determined by \cite{Udalski2018} is even less likely. In this case both components lie in the $3\sigma$ range.\\
\setcounter{figure}{3}  
\afterpage{%
\vspace*{\fill}
\begin{minipage}{\textwidth}
\centering
    \includegraphics[width=0.9\textwidth]{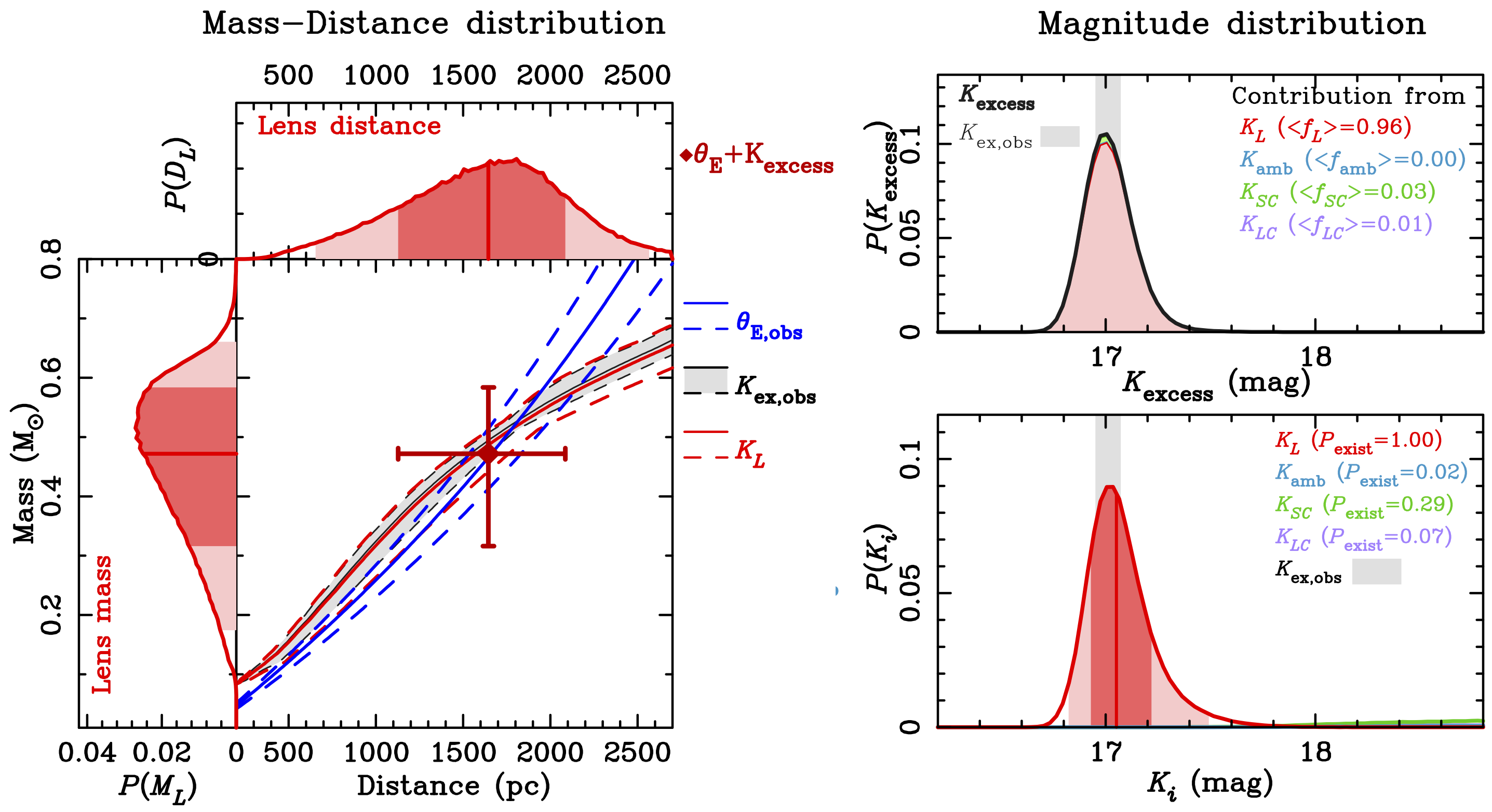}\\
    (a) Without parallax ($\pi_E$). Using constraints from $\theta_E$ and Keck excess flux ($K_{ex,obs}$) only.
    \includegraphics[width=0.9\textwidth]{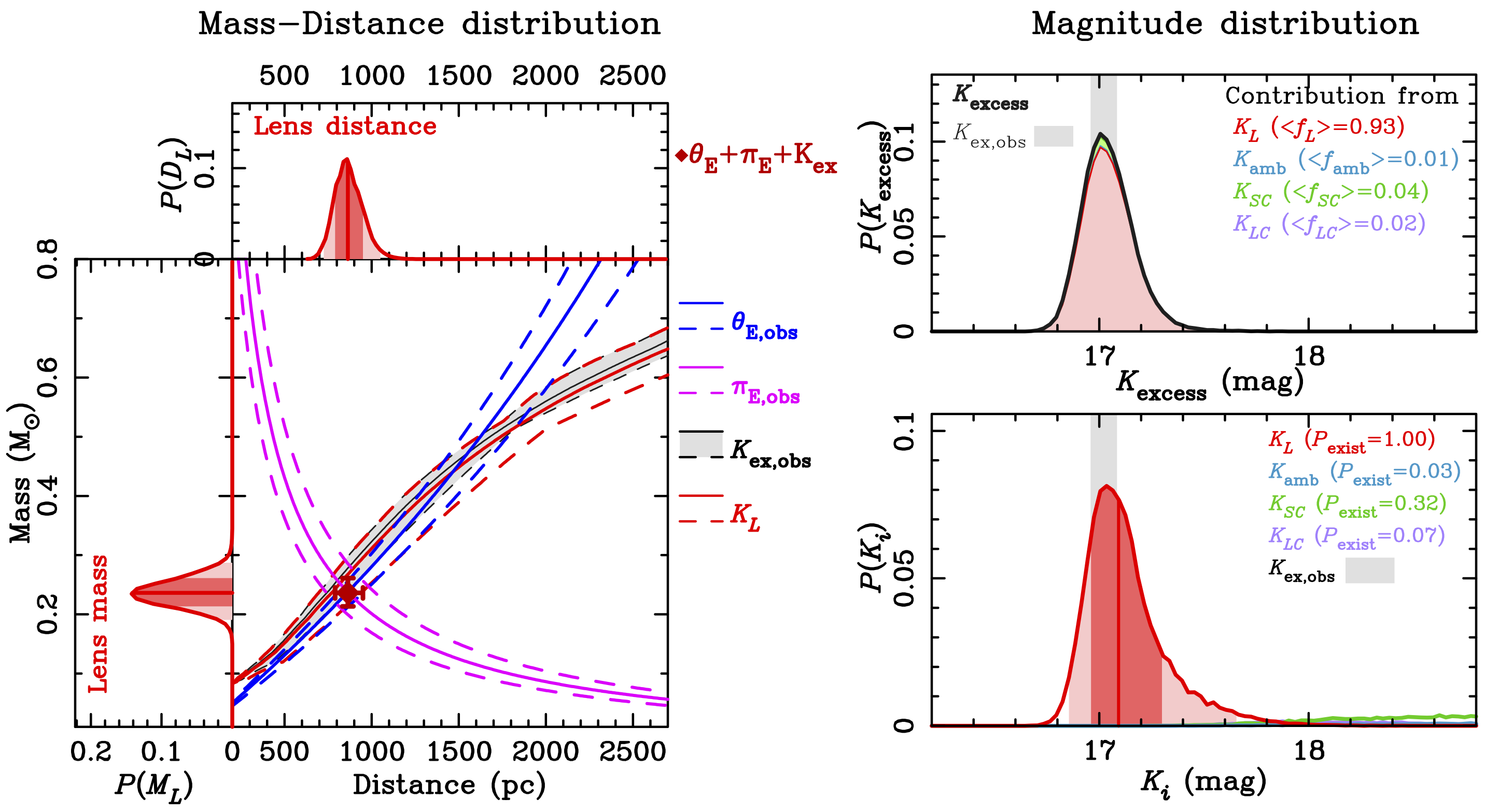}\\
    (b) With parallax ($\pi_E$). $\pi_E$, $\theta_E$ and Keck excess flux ($K_{ex,obs}$) constraint.
 \captionof{figure}{Posterior probability distributions showing the predicted mass-distance and magnitude distributions for OGLE-2017-BLG-1434. (a) The mass-distance distributions calculated using a Bayesian analysis with a galactic model prior and constraints from the measured Einstein ring radius ($\theta_{\mathrm{E,obs}}$) and Keck excess flux, ($K_{\mathrm{ex,obs}}$). The left panel shows 1, 2 and 3$\sigma$ histogram distributions of the lens mass and distance, with constraints shown for $\theta_{\mathrm{E,obs}}$ and $K_{\mathrm{ex,obs}}$. The right panel shows the probability of contributions from the lens ($K_L$), an ambient star ($K_{\mathrm{amb}}$), a companion to the source ($K_{SC}$) and a companion to the lens ($K_{LC}$). The upper right plot shows the total $K_{\mathrm{excess}}$ probability distribution, while the lower right shows the probability distributions of each component. In this case the excess flux is most likely from the lens ($K_L=0.96$). $P_{\mathrm{exist}}$ in the lower right panel indicates the probability of existence of each object. Figure (b) shows the distributions calculated in the same manner as (a), but with the added constraint from parallax}, $\boldsymbol{\pi}_{\mathrm{E,obs}}$. Note the values for $K_{\mathrm{amb}}$ and $K_{\mathrm{LC}}$ are too small to be visible on the right panels but are included here for completeness. Furthed details on the method used to create this figure can be found in \cite{Koshimoto2020}.
\label{fig:bayesian}
\end{minipage}
\vspace*{\fill}
\clearpage
}
\indent The parallax estimates are hence consistent with the large $\theta_E$ and our Keck measurements, but not very likely. These measurements rely on the measuring the microlensing parallax, $\boldsymbol{\pi}_{\mathrm{E}}$, most often from the orbital motion of the Earth \citep{Penny2016}. The component of parallax perpendicular to the Earth's motion ($\pi_{\mathrm{E},N}$), is degenerate with blending and the orbital motion of the lens and source (the effect on the light curve from these qualities mimics that from orbital parallax, meaning that they are difficult to disentangle from one another; \cite{Bennett2010a,Batista2011}). Caution, therefore, must be had when trusting $\boldsymbol{\pi}_{\mathrm{E}}$ measurements, in particular for close targets. A new reduction of the MOA data using the photometry code of \citep{Bond2017}, for example, indicates that the large parallax signal reported in the event MOA-2007-BLG-192 \citep{Bennett2008, Kubas2012} was incorrect. This was due to a photometric error due to color dependent differential refraction, first noted to be significant in \cite{Bennett2012}. \cite{Penny2016} meanwhile notes an unexpectedly large number of planets at close distances and cites six planetary events with $D_L < 2\:\si{kpc}$, of which two could be plausibly be moved to farther distances: OGLE-2013-BLG-0341 and OGLE-2013-BLG-0723. Since that publication OGLE-0723 \citep{Udalski2015,Han2016} has been revealed to be a binary event with no evidence of a planet, while variations in the baseline photometry of OGLE-0341 has been contradicted by data taken after \cite{Gould2014} was published, and whose large $\pi_E$ value could be caused by large systematic photometry errors. In the case of MOA-2007-BLG-192 \citep{Kubas2012}, the detection of a source excess using AO imaging is also consistent with an unrelated source contaminant, rather than solely the lens detection that was claimed.\\
\indent Our Bayesian analysis predicts the system to be a $m_p=4.43\pm 0.25\;M_\Earth$ planet orbiting a $M_L=0.234\pm 0.012\;M_\odot$ star at a distance of $0.86\pm0.05\;\si{kpc}$. The updated physical parameters of the lens and its companion calculated using this extra constraint are shown in Table \ref{tab:results}. The contaminant analysis of \cite{Koshimoto2020} employed here shows no tension between the reported parallax of \cite{Udalski2018} and our Keck measurements, however the values of parallax determined from the light-curve model (eg. $\boldsymbol{\pi}_{\mathrm{E}}= (\pi_{\mathrm{E},N},\pi_{\mathrm{E},E}) =(−0.586\pm0.081,0.472\pm0.013$ for a parallax+orbital motion fit where $u_0>0$) are unlikely and could be the results of systematic errors in the photometry. Excluding the parallax constraints we find that the excess flux at the position of the source is almost certainly the lens (with a probability of 0.96, Fig. \ref{fig:bayesian}, panel (a)). These probabilities were calculated using the $\theta_E$ and the Keck excess flux ($K_{ex,obs}$) constraint.  If we also include the parallax (Fig. \ref{fig:bayesian}, panel (b)) the excess flux is still very likely to be the lens ($K_L({<}f_L{>}=0.93$).\\
\section{Discussion \& Conclusion}\label{sec:conclusion}
We obtain Keck follow-up photometry of the  microlensing event OGLE-2017-BLG-1434 which is consistent with the physical interpretation of \cite{Udalski2018} that the system is a super-Earth planet orbiting an M4V dwarf star. When applying the additional constraint on the lens mass and distance from our lens flux measurement we reduced the uncertainty in the lens parameters (mass, distance, planetary mass) by half. As such we can now describe the system as a super Earth at a distance of $0.86 \pm 0.05\:\si{kpc}$,. This is quite nearby for a microlensing planet but farther away than the majority of radial velocity and transit-detected planets.  The planet and its host star had an instantaneous 2D separation of $a_\perp=1.18 \pm 0.10$, which places a lower limit on its perihelion distance. Calculating this using the relationship $a_\perp=s\theta_E D_L$ results in a value in agreement with \cite{Udalski2018}, but with an error reduced by 28\%. Comparing this to the snow line, defined as $a_{snow} = 2.7\si{AU}\;(M/M_\odot)$, the planet lies at $1.9 \pm 0.2\;\mathrm{a_{snow}}$, while the stellar lens host has a mass of $0.234\pm0.012\:\si{M_\odot}$.\\
\subsection{Galactic Environment}

\indent This planet joins a list of only 5 other planets detected by microlensing at a distance of less than 1 kpc, assuming no systematic errors in the photometry (see Section \ref{sec:bayes}). Of the 100 or so planets so far detected using microlensing, the median distance of these systems, almost all in the direction of the galactic centre, is $\sim5.7\;\si{kpc}$\footnote{https://exoplanetarchive.ipac.caltech.edu/}. \\
\indent One consequence of this distance distribution is that the typical microlensing-detected planet host is expected to be significantly more metal-rich than planetary host stars in the Solar neighborhood, owing to the radial metallicity gradient in the Galactic disk. The inner galaxy (R$_{GC}$ $\la$ 3~kpc) is expected to contain the most metal-rich stars in the Milky Way \citep[e.g.,][]{Larson1976}; surveys have found this to be the case \citep[e.g.,][and references therein]{Zoccali2008}, with the kinematically cold population near the disk plane home to proportionally more stars of super-Solar metallicity, reaching up to [Fe/H] $\sim +0.4-0.6$ \citep{McWilliam1994,Ness2013}. \\
\indent Thin disk stars within $\approx$3--5~kpc of the Galactic centre have a relatively narrow peak in their metallicity distribution at [Fe/H] $\approx$ $+$0.3 \citep[e.g.,][]{Hayden2015}, with a tail to lower values and little evidence for radial gradients \citep{Hayden2014}. By contrast, in the Solar neighborhood the radial metallicity gradient amounts to $-$0.09 dex/kpc \citep{Frinchaboy2013,Hayden2014}. The mean metallicity of dwarfs and subgiants in the Solar neighborhood is [Fe/H] = $-$0.04, with a dispersion of $\sigma$ = 0.25~dex and a roughly Gaussian shape \citep{Buder2019}.\\
\indent The stellar population structure of the Galactic disk leads us to expect a metallicity difference of a factor of 1.5--2 between typical planetary hosts in the inner Galaxy and the Solar neighborhood, which must be reckoned with when comparing planet occurrence as a function of host mass and orbital period \citep[e.g.][]{Sousa2019} across samples from different Galactic environments. For microlensing surveys, continuing to build exoplanet samples at a wide range of distances will allow for a statistical exploration of the frequency of cold planets as a function of host star metallicity. This is highly complementary to transit and Doppler exoplanet searches in the Solar neighborhood, which are accessible to direct spectroscopic probes of stellar metallicity but are biased towards the detection of hot planets around more intrinsically luminous stars.

\subsection{Mass-Ratio Function}

\indent OGLE-2017-BLG-1434 is one of five microlensing planets with a mass ratio of $ q \leq 0.6 \times 10^{-4}$ and one of 11 planets with a mass between $1-10M_\Earth$. It lies right on the break of the exoplanet mass-ratio power-law according to \cite{Jung2019}, which is smooth and decreasing at mass ratios higher than this inflection point, and increasing below. The slope at these smaller mass ratios is not well constrained with \cite{Jung2019} claiming a sharp break at the inflection point of the mass-ratio function. They give a best fit value of $q = 0.55 \times 10^{-4}$. The lack of statistics in this region is a significant limitation in attempts to determine the slope and shape of the low-$q$ mass-ratio function. This is why lens flux measurements such as those presented here are important. Even though it requires high-resolution imaging from the largest ground based telescopes, or from the Hubble Space Telescope, the time costs for each event is small (usually 30-40 mins excluding calibration overhead).\\
\indent Of the eleven planets with mass ratios $q < 1 \times 10^{-4}$, four of them have reliable host mass measurements: OGLE-2005-BLG-169Lb \citep{Batista2015}, MOA-2009-BLG-266Lb \citep{Muraki2011}, KMT-2018-BLG-0029Lb \citep{Gould2020} and OGLE-2017-BLG-1434 (this study), while a number of others have estimates computed using a Bayesian analysis that are highly dependent on the priors: OGLE-2005-BLG-390Lb \citep{Beaulieu2006,Kubas2008}, OGLE-2007-BLG-368Lb \citep{Sumi2010a}, OGLE-2017-BLG-0173Lb \citep{Hwang2017}, OGLE-2018-0677Lb \citep{Herrera-Martin2020} and KMT-2019-BLG-0842Lb \citep{Jung2020}. 
Especially for events with limited secondary light-curve effects and physical parameters only determined to 30-40\%, single band near infrared follow-up photometry as shown here is a time-cheap way of tightening mass and distance constraints on microlensing systems, and -- eventually -- of characterising the break in the planetary mass-ratio function.\\

\indent This work was supported by the University of Tasmania through the UTAS Foundation and the endowed Warren Chair in Astronomy. It was also supported by ANR COLD-WORLDS (ANR-18-CE31-0002) at Le Centre National de la Recherche Scientifique (CNRS) in Paris and the Laboratoire d'astrophysique de Bordeaux. AAC and JWB are supported by funding from the Australian Research Council through the Discovery Project grant scheme (DP200101909). DPB and AB were supported by NASA through grant NASA-80NSSC18K0274. Data presented in this work was obtained at the W. M. Keck Observatory from telescope time allocated to the National Aeronautics and Space Administration through the agencies scientific partnership with the California Institute of Technology and the University of California. These Keck Telescope observations and analysis were supported by a NASA Keck PI Data Award, administered by the NASA Exoplanet Science Institute. The Observatory was made possible by the generous financial support of the W. M. Keck Foundation. This research has made use of the NASA Exoplanet Archive, which is operated by the California Institute of Technology, under contract with the National Aeronautics and Space Administration under the Exoplanet Exploration Program.



\bibliographystyle{yahapj}
\bibliography{library}

\end{document}